# A short history of Quantum Illumination

Marco Genovese[a], Ivano Ruo-Berchera[a]
[a]National Metrology Institute of Italy (INRIM), *strada delle Cacce 91, 1035 Turin, Italy*

**ABSTRACT**

Quantum illumination represents one of the most interesting examples of quantum technologies. On the one hand, it can find significant applications; on the other hand, it is one of the few quantum protocols robust against noise and losses.

Here we present a short summary of the history of this quantum protocol.

## 1. INTRODUCTION

In the last 3 decades, the possibility of manipulating single quantum systems, as single photons or single ions, led to the development of technologies based on the peculiar properties of quantum states, as entanglement [1].

The first quantum technology that was developed was quantum information, studying the possibility of coding, elaborating (quantum computation [2]) and transmitting (quantum communication [3]) information.

A little later, it was suggested to exploit peculiar properties of quantum states for improving measurement capabilities, the so-called quantum metrology [4-7] and sensing [8,9], that already led to significant applications ranging from detecting magnetic fields on nanoscale [10] to measuring temperature variation in neurons when action potential is excited [11].

In this area, in the optical domain quantum imaging & sensing emerged as a discipline with huge opportunities for practical applications [12].

Several different schemes have been proposed and already realized experimentally. Among them it is worth mentioning ghost imaging (GI) [12-17], quantum lithography [18,19], entangled images [20], image amplification by Parametric Down Conversion (PDC) [21,22], quantum enhanced interference microscope [23], super resolution by photon statistics evaluation [24-30], imaging with undetected photons [31], infrared spectroscopy with visible light [32], quantum reading and pattern recognition [33-35] and last but not least, being the first truly quantum imaging experiments, in the sense of not being realisable with classical light as GI, sub shot noise imaging [36-39].

## 2  QUANTUM ILLUMINATION

A scheme of particular interest, that can find widespread applications ranging from civil surveillance, environmental monitoring and pollution detection to defense, is Quantum Illumination (QI), Fig.1. Differently from most of the other quantum technologies, QI offers a significant advantage in detecting an object in a preponderant noise. The QI is a revolutionary concept at the intersection of quantum physics and engineering, promising to transform how we detect and image objects in environments where classical methods fall short. Unlike traditional radar or LIDAR, which rely on classical electromagnetic waves, quantum illumination exploits the unique properties of quantum entanglement and quantum states of light to achieve superior detection sensitivity, especially in noisy or low-visibility conditions. This technology has far-reaching implications for fields as diverse as medical imaging, underwater communication, and stealth object detection and covert sensing.

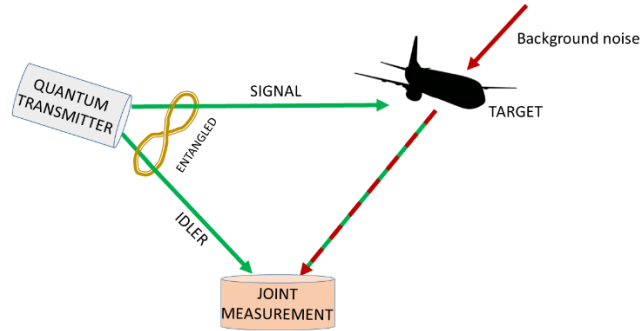

*Figure 1: Quantum Illumination Scheme*

In particular, this quantum protocol addresses the problem of detecting an object by illuminating it with quantum light, even if the light reflected back is much smaller than the background noise. At the heart of QI lies the phenomenon of quantum correlations. The QI scheme is illustrated in Figure 1. A pair of correlated photons is generated: one, called the "signal" photon, is sent toward the target object, while the other, the "idler" photon, is retained for measurement.

When the signal photon interacts with the target and is reflected back, it carries information about the object's presence and properties. The idler photon, which has never interacted with the target, is used in conjunction with the returned signal photon to perform a joint measurement. This process leverages quantum correlations to distinguish the reflected signal from background noise, significantly enhancing detection sensitivity. Classical detection methods, such as radar, suffer from limitations in noisy environments because the reflected signal can be overwhelmed by background radiation from the environment or intentionally added to jamming purposes. QI, however, uses the quantum correlation between the signal and idler photons to filter out noise, allowing for the detection of objects that would otherwise be invisible to classical sensors.

One of the most promising applications of QI is in the detection of stealth objects, such as low-observable aircraft or submarines. Traditional radar systems struggle to detect these objects due to their ability to absorb or scatter radar waves, where only a small signal is reflected back. QI, by exploiting quantum correlations, can potentially detect such objects even when they are designed to evade classical sensors. Another field of interest is represented by underwater environments, which are notoriously challenging for classical imaging and communication technologies due to the strong absorption and scattering of light. QI offers a way to overcome these challenges by using entangled photons to transmit information with higher fidelity and lower error rates, enabling more reliable underwater communication and imaging.

Also, in the medical field, QI could revolutionize imaging techniques by providing higher resolution and lower noise images. For example, quantum-enhanced imaging could improve the detection of tumours or other abnormalities in tissues, leading to earlier and more accurate diagnoses.

Following some idea emerged about the distinguishability of entanglement breaking channels [40], QI was proposed theoretically in 2008 [41]. The main idea of this paper was to demonstrate, by exploiting the Chernoff bound, that quantum properties of light, specifically entanglement, can be used for increasing exponentially the probability of detecting an object in a preponderant noise when looking at the signal reflected back.

This first proposal was based on disposing of a perfect source of high-dimension single photon entangled state, of a lossless storage of one of the two components and of a perfect receiver. This result was very exciting and attracted large interest, nonetheless no real scheme was proposed for its realization.

A step in this sense was achieved by Ref. [42], in which it was demonstrated that a Gaussian state, e.g. the "realizable" twin beams, with a phase-sensitive measurement allows a 6 dB advantage over classical illumination.

Following these theoretical proposal, the first experimental realization of a QI protocol was then realized for the first time in INRIM in the early 2012 [43]. The specific purpose of this experiment was to demonstrate QI, in the general meaning of using of quantum light for target detection, choosing a configuration suited for a practical application. In this sense, twin beams were used, as suggested in [42], while the detection scheme was simply based on coincidence detection. It was demonstrated that this allows for an exponential advantage with respect to a classical scheme when one wishes to detect an unknown object in a preponderant unknown background when the signal is faint.

After this first realization, several other experiments extended this result in the optical domain [44-46], by adopting an interesting detection scheme, albeit sub-optimal, realized by combining the return and idler modes in an optical parametric amplifier; the method was applied to secure quantum communication.

Furthermore, on the one hand, the resources relevant for QI were investigated [47,48] and extensions toward regions where a particular low dose is necessary, as X-rays [49], or a very strong background is present, as microwaves [50-53], were studied and realized. Typically, in microwave domain, the entanglement generation is achieved through Josephson Parametric amplifier.

These last studies are particularly interesting in order to move toward the realisation of "true" quantum radar and LIDAR, i.e. where not only an object is detected but also its distance is determined [54-58]. Very recently, it has been shown that in certain practical contexts, the concept of QI applied to LIDAR allows covert sensing, i.e. performing target detection without being discovered [59], while this is not possible with a classical transmitter.

# 3 CONCLUSIONS

In conclusion, quantum correlations of light, after having played a major role in studies on foundations of quantum mechanics [1,60], are now a fundamental tool for developing quantum technologies, as well as for enhancing the performances of imaging, allowing to beat classical limits, such as shot noise, or realising innovative schemes [61,62]. A very important aspect of these protocols is that several of them are based on resources realisable easily in the laboratory, such as twin beams [63-65], and, therefore, are ready for practical, even commercial, applications. Among these, quantum illumination, and its extensions to quantum LIDAR and quantum radar, are among the most significant and not far from practical application [66,67]. Indeed, quantum illumination represents a paradigm shift in detection and imaging technology. By harnessing the power of quantum entanglement, it offers unparalleled sensitivity and noise resistance, opening up new possibilities in fields ranging from defense to medicine. While significant challenges remain, ongoing research and technological advancements are steadily bringing quantum illumination closer to practical realization.

# ACKNOWLEDGEMENTS

This paper has received funding from the project Nato-G5263 and from Foundation San Paolo, project QuteNoise